\documentclass[twocolumn]{revtex4}  
  
\usepackage[utf8]{inputenc}
\usepackage{graphicx}
\usepackage{amssymb}
\usepackage{amsmath}
\usepackage{xcolor}
\usepackage{soul}

\usepackage[mathlines]{lineno}
\usepackage[normalem]{ulem}

\begin{document}

\title{
Efficiency of random search with space-dependent diffusivity 
}

\author{M. A. F. dos Santos$^{1}$,  L. Menon Jr.$^{1}$,  and C. Anteneodo$^{1,2}$}
\address{$^{1}$ Department of Physics, PUC-Rio, Rua Marqu\^es de S\~ao Vicente 225, 22451-900, Rio de Janeiro, RJ, Brazil}
\address{$^{2}$ Institute of Science and Technology for Complex Systems, INCT-SC, Brazil}

\begin{abstract}
We address the problem of  random search for a target in an environment with space-dependent diffusion coefficient $D(x)$. From a general form of the diffusion differential operator that includes  Itô, Stratonovich, and Hänggi-Klimontovich interpretations of the associated stochastic process, 
we obtain the first-passage  time distribution and the search efficiency $\mathcal{E}=\langle 1/t \rangle$. For the paradigmatic  power-law  diffusion coefficient $D(x) = D_0|x|^{\alpha}$, with $\alpha<2$, which controls whether the mobility increases or decreases with the distance from a target at the origin, we show the impact of the different interpretations. 
For the Stratonovich framework, we obtain a closed expression of the search efficiency, valid for   arbitrary diffusion coefficient $D(x)$.    %
We show that a heterogeneous diffusivity profile leads to lower efficiency than the homogeneous  average level, and the efficiency   depends only  on the distribution of diffusivity values and not on its spatial organization, 
 features that breakdown under other interpretations.  
 
\end{abstract}


\maketitle


\section{Introduction} \label{sec:intro}

In many complex environments, the diffusivity cannot be considered uniform, but  changes from one point to another~\cite{cherstvy2013anomalous,nature2013}. 
State-dependent diffusivity  has been considered to describe  particles moving between nearly parallel plates~\cite{lanccon2001drift},
biologically motivated problems~\cite{pieprzyk2016spatially,berezhkovskii2017communication,dos2020critical,dos2021random},  and stock markets ~\cite{oksendal2013stochastic}, among many others. Recently, heterogeneous diffusion processes (HDP) have been also investigated within the stochastic resetting scenario ~\cite{sandev2022heterogeneous}. 
In one-dimension, a single trajectory $x(t)$ can be modeled by the following stochastic process 
\begin{eqnarray} \label{eq:langevin}
\dot{x} =  \sqrt{2D(x)}\,\eta(t),
\end{eqnarray}
where $x$ is the spatial coordinate (or other state variable, such as chemical coordinate or stock prize), $D(x)>0$ is the diffusion coefficient,  $\eta(t)$ is a zero-mean white noise with  delta-correlation $\langle \eta(t+t')\eta(t)\rangle = \delta(t')$. 
Due to the white noise and multiplicative character of the stochastic   Eq.~(\ref{eq:langevin}), its integration requires an  additional specification~\cite{risken1996fokker}. 
In any case, the stochastic Eq.~(\ref{eq:langevin}) can be cast, for instance, in the It\^o form, appropriate for numerical simulations, by  adding  a drift term, namely,
 \begin{eqnarray} \label{eq:ito}
\dot{x}  =  (1- A/2 ) \frac{d D(x)}{dx} + \sqrt{2 D(x)}\,\eta(t)\,, 
 \end{eqnarray}
 where $A$ controls the interpretation~\cite{ito1944109,stratonovich1966new,hanggi1982nonlinear,nature2013}. Standard cases are
 $A=0$ (Hänggi-Klimontovich or isothermal),   $A=1$ (Stratonovich), and 
 $A=2$ (Itô, in which case the drift term vanishes), although other values of $A$ have also been  considered~\cite{hottovy2012noise}. For homogeneous diffusivity, the drift vanishes.

 \begin{figure}[b!]  
\centering
 \includegraphics[width=0.95\columnwidth]{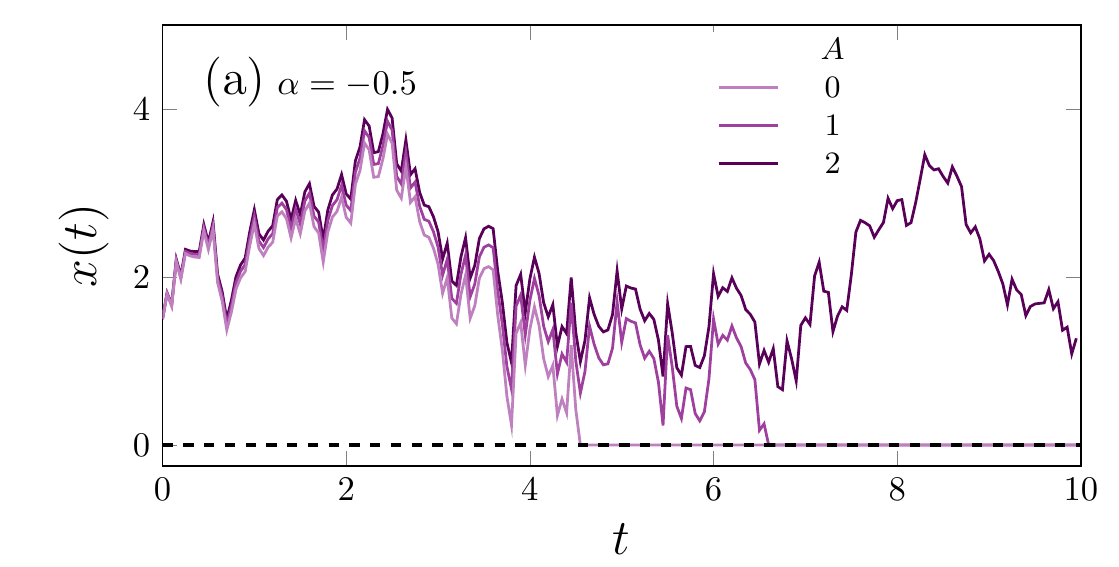} 
\includegraphics[width=0.95\columnwidth]{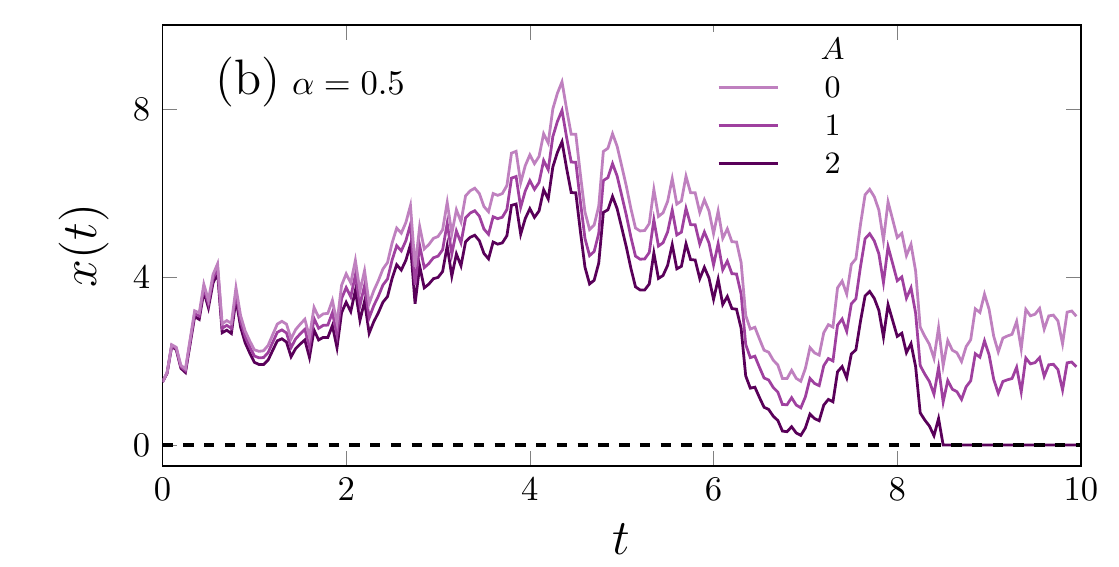}
\caption{  Random trajectories in a heterogeneous environment  with $D(x) = D_0 x^{\alpha}$ and an absorbing boundary at $x=0$ (dashed line), for  $\alpha = -0.5$ (a) and $\alpha=0.5$ (b). 
We set  $D_0 = 1.0$, $x_0 = 1.5$, and the integration was performed using the stochastic Euler algorithm with time step $dt=0.05$. In each panel, the same random sequence was used, for comparison.
} 
\label{fig:trajectories} 
\end{figure}

The corresponding heterogeneous diffusion equation is
\begin{eqnarray}
 \partial_t p(x,t) =   \partial_x 
\left\{ D(x)^{1-A/2}  
 \partial _x 
\left(D(x)^{A/2} p(x,t) \right) 
  \right\},\;\;\;
\label{eq:HDP}
\end{eqnarray}
where $p(x,t)$ is the probability density function (PDF). The parameter $A$ has impact on the spreading of particles, and can change the tails of the PDF, produce diffusion anomalies and ergodicity breaking~\cite{PhysRevE.99.042138,cherstvy2013anomalous,bressloff2017temporal,dos2018fractional}.

In Fig.~\ref{fig:trajectories}, we show typical trajectories when the diffusivity has the power-law form $D(x)=D_0x^\alpha$,  for $\alpha=\pm 0.5$ and an absorbing boundary at $x=0$, for three different values of $A$.  For each interpretation, the stochastic term is the same (and we used the same random sequence of the noise  for comparison) but the deterministic drift term is enhanced with decreasing $A$. Moreover, it is either positive (if $\alpha>0$) or negative (if $\alpha<0$).  Then, increasing $A$ will make the walker reach the origin for the first time earlier or later, respectively, as observed in each panel of the figure.

Therefore, the details of a heterogeneous environment are expected to have important consequences in random searches problems~\cite{redner2001guide,risken1996fokker}, 
a class of problems that is relevant in diverse contexts and at different scales~\cite{benichou2011intermittent,zaburdaev2015levy}.  
At the molecular level, let us mention the search of a protein for its binding site on DNA~\cite{mirny2009protein,chen2019target,bhattacherjee2014search}, 
at the ecological scales,  the search for 
food (foraging)~\cite{o1990search,viswanathan1999optimizing,viswanathan2001levy,bartumeus2005animal,viswanathan2011physics}. 
Other applications  include design in robotics~\cite{castello2016adaptive} or  computer algorithms to search minima  in a complex landscape~\cite{pavlyukevich2007levy}. 
In all these cases, finding efficient strategies that minimize the time to encounter the target, or optimize other search criteria, is crucial.  
In this context, several diffusion processes have been investigated, for instance, L\'evy flights \cite{palyulin2017comparison}, fractional Brownian motion \cite{khadem2021search}, Brownian search in quenched heterogeneous media \cite{godec2015optimization}, run-and-tumble \cite{rupprecht2016optimal}, and resetting ~\cite{chechkin2018random,bhat2016stochastic}. 
Search in heterogeneous diffusivity media has been studied  
for particular forms of the diffusion coefficient in a confined setting, for Hänggi-Klimontovich interpretation (theoretically)~\cite{godec2015optimization},
for Stratonovich (numerically)~\cite{mutothya2021first},
or with stochastic resetting~\cite{lenzi2022transient,Ray2020}.
The step shape of the diffusivity profile in a confined and $d$-dimensional system was investigated in Ref.~\cite{vaccario2015first} for all interpretations.
In all these cases the  mean first passage time (MFPT) was calculated, but the average of the inverse time (efficiency) is another relevant quantity, that has been calculated for instance for 
random search on a comb model~\cite{sandev2020hitting}.
Our purpose is to measure the search efficiency in nonconfined heterogeneous media with arbitrary diffusivity profile and arbitrary interpretation.

A target can be introduced into the diffusion equation by means of a  $\delta$-delta sink term or by  absorbing boundary conditions. We will use the latter approach. 
Moreover, we consider that a searcher follows a  HDP, exploring all points along its trajectory~\cite{benichou2011intermittent}.  
To study random searches, 
it is central  to determine the first-passage-time distribution (FPTD)  
\begin{eqnarray} \label{eq:wp}
\wp(t)=-\frac{d\ }{dt} Q(x_0,t)\,,
\end{eqnarray}
where $ Q(x_0,t) = \int_{\Omega}p(x,t|x_0)dx$, with support  $\Omega$ is the survival probability at time $t$. The FPTD represents the probability density of the first time  the walker meets the target, after which the walker is removed from the system~\cite{redner2001guide,risken1996fokker}. 
Therefore, note that  the norm of the density $p(x,t|x_0)$ is not conserved.

To quantify and compare the performance of different search processes, a fundamental measure  is the so called search efficiency, and various definitions  can be found in Ref.~\cite{James2010}. 
We will use a definition close to the  step efficiency (inverse of the  traveled time up to reaching the target)~\cite{palyulin2014levy}, namely, 
\begin{eqnarray}
\mathcal{E} = \left \langle t^{-1} \right \rangle = \int_0^{\infty} t^{-1}\, \wp(t) dt  = \int_0^{\infty} \widetilde{\wp}(s) ds
,
\label{eq:efficiency}
\end{eqnarray}
where $\widetilde{\wp}(s)$ is the Laplace transform of the FPTD. The measure  defined by Eq.~(\ref{eq:efficiency}) is adequate for systems where the mean arrival time diverges~\cite{palyulin2014levy}. 
$\mathcal{E}$ is the first-order negative moment, which straightens the contribution of short arrival times.  
It has been used in a series of works  to characterize the performance of L\'evy searches, facing multiples targets~\cite{palyulin2017comparison}, under external bias~\cite{Palyulin2014bias}, 
comb structures~\cite{Sandev2019}, asymmetric Lévy flights  \cite{padash2022asymmetric} and to describe long relocations mingled with thorough local exploration \cite{palyulin2016search}.

Our results are organized as follows. 
In Sec.~\ref{sec:plaw},  we use  the backward Fokker-Planck equation, with arbitrary $A$, to obtain the first  passage time distribution and the search efficiency when the position-dependent diffusivity has a power-law form, which has been used in different frameworks \cite{cherstvy2013anomalous,sandev2022heterogeneous,sandev2018heterogeneous}.
In Sec.~\ref{sec:stratonovich}, we obtain a closed expression for the efficiency, valid for arbitrary $D(x)$, when the prescription is of Stratonovich type ($A=1$).   
In all cases, examples and comparisons of the theory with stochastic simulations are provided. 
Final remarks are presented in Sec.~\ref{sec:final}.

\section{Random search in media with power-law diffusivity under different prescriptions}
\label{sec:plaw}

\subsection{Survival probability}

We consider independent random walkers on a one-dimensional heterogeneous medium,  initially located at position $x_0$, i.e., the initial density function is $p(x,0)=\delta(x-x_0)$ and its  evolution   is described by Eq. (\ref{eq:HDP}), which can be rewritten as
\begin{eqnarray} \nonumber
\frac{\partial \ }{\partial t} p(x,t|x_0) &= &    \frac{\partial^2 \ }{\partial x^2} \left\{D(x)p(x,t|x_0)  \right\} \\
&+& \left(A/2-1\right) \frac{\partial }{\partial x} \left\{\frac{dD(x)}{dx \ }p(x,t|x_0)  \right\}. \label{eq:ItoInterpretation}
\end{eqnarray}
In this format, the diffusion term is of the  It\^o form but a spurious drift term appears, which vanishes for $A=2$. This representation will be useful to obtain the survival probability.

To address the random search problem, we consider,  without loss of generality, that a target is located at $x= 0$, which corresponds to a change of coordinate. 
The target position  defines a bound of the search domain, because we are considering a cruise search in which
 the walker can detect the target during its movement, and it is removed when the target is first detected. 
 Without loss of generality, we assume that the  initial position of the random searcher is $x_0>0$, in which case the search domain is the positive $x$-axis $[0,\infty)$. 

Regarding the heterogeneous diffusivity, in this section we will focus our analyses on the power-law case 
\begin{equation} \label{eq:Dx}
    D(x)=D_0\,x^{\alpha}\,,
\end{equation}
where $x\ge 0$ and $\alpha<2$. 
This kind of profile has been used to capture the diffusive motion of a particle on   fractal objects~\cite{PhysRevLett.54.455} and   
diffusion in turbulent media \cite{sandev2020hitting}. 
It has also been  used 
as a paradigm of heterogeneous diffusivity to study infinite ergodic theory~\cite{PhysRevE.99.042138}, extreme value statistics ~\cite{PhysRevE.105.024113}, and critical habitat size of biological populations~\cite{dos2020critical}. 
In the current problem,  we can interpret that the target modifies the mobility of the searcher around it, making it increase ($\alpha>0$) or decrease ($\alpha<0$)    with distance.  

 The survival probability  $Q(x_0,t)=\int_{0}^{\infty}p(x,t|x_0)dx$  represents the probability that the diffusing particle, starting at $x_0$, has not hit the target  ($x=0$) up to time $t$. 
 It can be determined through the backward Fokker-Planck equation~\cite{risken1996fokker}, which, for the chosen power-law $D(x)$, reads
\begin{eqnarray} \nonumber
\frac{\partial \ }{\partial t} Q(x_0,t) & = &    D_0 x_0^{\alpha} \frac{\partial^2 \ }{\partial x_0^2} Q(x_0,t) \\
&+&\frac{\left(1-A/2\right) D_0 \alpha} {x_0^{1-\alpha}} \frac{\partial }{\partial x_0}  Q(x_0,t) ,\;\;\;\;\;\; \label{eq:backward}
\end{eqnarray}
together with the boundary condition $Q(0,t)=0$, meaning that the survival has null probability when the walker  starts at  the target position $x_0=0$. 
The initial condition is $Q(x_0>0,0)=1$, since 
$p(x,0|x_0)= \delta(x-x_0)$ and $x_0 \in (0,\infty)$.

The Laplace transform, defined as $\Tilde{f}(s) = \int_0^{\infty}f(t)e^{-st}dt$, applied on the time variable of Eq.~(\ref{eq:backward}), implies
\begin{eqnarray} \nonumber
s\Tilde{Q}(x_0,s) - 1 & = &   D_0  x_0^{\alpha} \frac{\partial^2 \ }{\partial x_0^2} \Tilde{Q}(x_0,s) 
\\ &+&  \frac{\left(1-A/2\right) D_0  \alpha}{x_0^{1-\alpha }}  \frac{\partial }{\partial x_0}  \Tilde{Q}(x_0,s) , \;\;\;\;\;\; \label{eq:surveq}
\end{eqnarray}
which can be solved analytically, as shown in Appendix~\ref{app:solution}, obtaining
\begin{eqnarray} \nonumber
\Tilde{Q}(x_0,s) & = &  \frac{1}{s} -  \frac{2  }{   \Gamma[b]\,s}\left( \frac{x_0^\frac{2-\alpha }{2}}{2-\alpha} \sqrt{\frac{s}{D_0}} \right)^{b}  \times \\
&\times& K_{b} \left( \frac{2\, x_0^\frac{2-\alpha }{2} }{2-\alpha} \sqrt{\frac{s}{D_0}}  \right)\,,
\label{eq:FinalQLaplace}
\end{eqnarray}
where $K_b(z)$ is the modified Bessel function
and 
\begin{equation} \label{eq:b}
    b= \frac{1-\alpha(1-A/2)}{ 2-\alpha }\,,
\end{equation}
with the restriction
 $1-\alpha(1-A/2) \ge 0$.
  See Eq. (\ref{eq:constraint}) in the appendix \ref{app:solution}.

\subsection{First passage time distribution}

From Eq.~(\ref{eq:wp}), the Laplace transform of the FPTD  is given by $\Tilde{\wp}(s)=1-s\Tilde{Q}(x_0,s)$, then
\begin{eqnarray} \label{eq:wps}
\Tilde{\wp}(s) =  \left( \frac{x_0^\frac{2-\alpha }{2} }{2-\alpha} \sqrt{\frac{s}{D_0}} \right)^{b} \frac{2}{ \Gamma[b]} \,K_{b} \left( \frac{2\,x_0^{\frac{2-\alpha }{2}}}{2-\alpha} \sqrt{\frac{s}{D_0}}  \right).
\end{eqnarray}
To perform Laplace inversion,  we use  
$\mathcal{L}^{-1}\left\{ s^{\frac{b}{2}} K_{b}\left( d \sqrt{s}  \right) \right\} =
\exp(-  d^2/[ 4 t])\, d^{b} /(2 t)^{b+1}$~\cite{abramowitz1965handbook}, 
which implies  
\begin{eqnarray}
\wp(t) =  \frac{1  }{\mathcal{Z} (2t)^{b+1}} \,e^{ -   \displaystyle{\frac{x_0^{2-\alpha }}{(2-\alpha)^2D_0\, t} } }\,, \label{eq:FPT}
\end{eqnarray}
with 
\begin{equation}
\mathcal{Z}^{-1}= 2\left( \frac{2}{2-\alpha}  \right)^{2b}  \frac{  x_0^{1  - \alpha \left(1-A/2 \right)}}{(2 D_0)^{b} \Gamma(b) }   ,
\end{equation}
recalling that, from Eq.~(\ref{eq:b}), it must be $\alpha\left( 1 - A/2 \right)<1$, for the FPTD to be normalizable.  In the particular case,  $A=2$ (It\^o interpretation) and $\alpha=1$, Eq.~(\ref{eq:FPT})
gives $\wp(t) = (x_0/D_0) t^{-2} \exp[-x_0/(D_0 t)]$, recovering previous results~\cite{Ray2020}.

 \begin{figure}[b!]  
\centering
 \includegraphics[width=0.95\columnwidth]{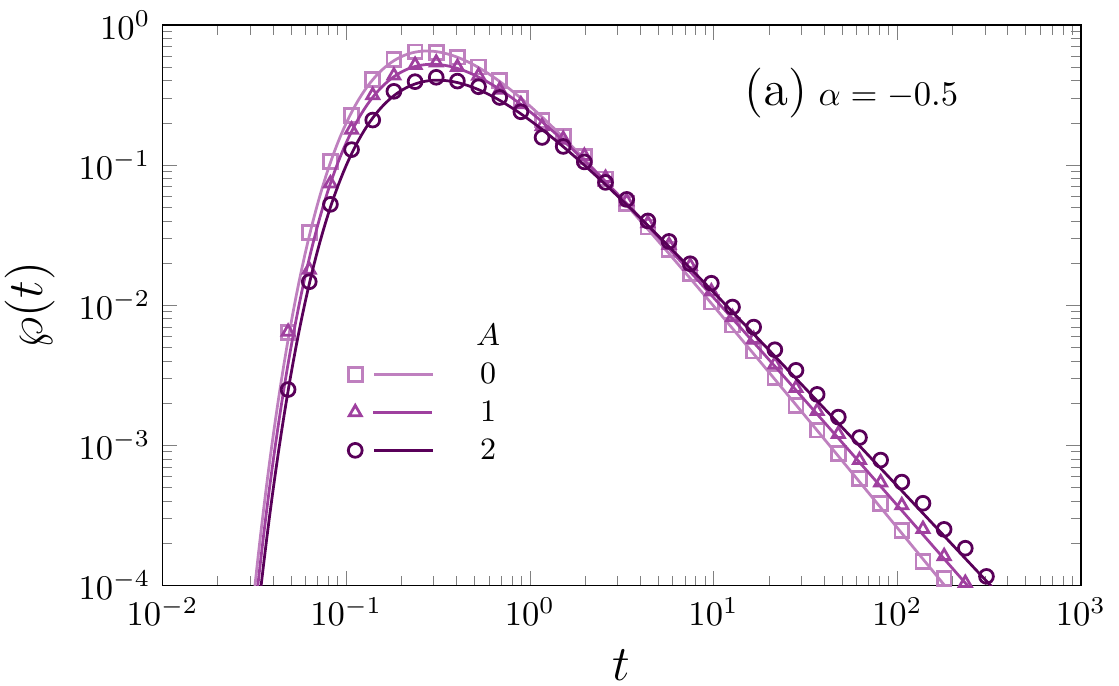}
 \includegraphics[width=0.95\columnwidth]{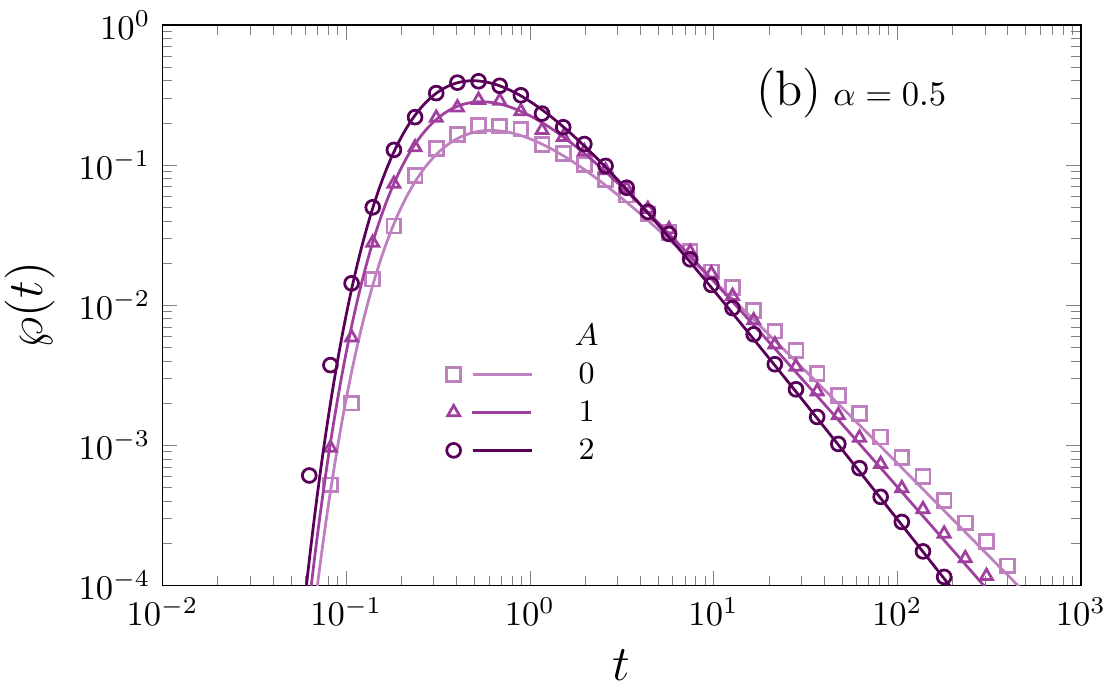}
\caption{First passage time distribution (FPTD) given by  Eq.~(\ref{eq:FPT})  (solid lines) and from $10^5$ trajectories obtained from    Eq.~(\ref{eq:ito}) with absorbing boundary at $x=0$ (symbols), for the HDP with $D(x)=D_0 x^\alpha$. Panels (a) and (b) are for $\alpha = -0.5$ and $\alpha = 0.5$, respectively. In all cases  $x_0 = 1.0$ and $D_0 = 1.0$. In simulations, for $t<1$ we used $dt=10^{-4}$, and for $t\ge1$ we used $dt=10^{-2}$. 
} 
\label{fig:FPT1}
\end{figure}

Figure~\ref{fig:FPT1} shows the good correspondence between the obtained FPTD from the analytical prediction given by  Eq.~(\ref{eq:FPT}) and from simulations of Eq.~(\ref{eq:ito}) with absorbing wall at $x=0$, for different values of $A$. 
Notice, for instance in case $\alpha=-0.5$ (a) that increasing $A$ diminishes the probability of short times and produces longer tails, two features that contribute to the tendency shown in Fig.~\ref{fig:trajectories}, delaying the encounter of the walker with the absorbing wall. The opposite occurs in case $\alpha= 0.5$ (b), also in accord with Fig.~\ref{fig:trajectories}.

Moreover, the asymptotic behavior of $\wp(t)$ in Eq.~(\ref{eq:FPT}) 
 indicates that normalization is possible when  $\alpha<2$ ($A=1,2$) or
$\alpha<1$ ($A=0$), which also implies the existence of the efficiency.
The mean first passage time $\langle t\rangle$ is finite only for $A=2$ and $1<\alpha<2$.

\subsection{Search efficiency} 
 
The search efficiency, defined in Eq.~(\ref{eq:efficiency}), can be calculated  using Eq.~(\ref{eq:wps}), which gives
\begin{eqnarray} \nonumber
\mathcal{E} 
    & = & 
    \frac{D_0}{x_0^{2-\alpha}}\left(1-\frac{2-A}{2}\alpha\right)(2-\alpha) \,.
    \label{eq:ef-alfa}
\end{eqnarray}

 \begin{figure}[b!]  
 \includegraphics[width=0.95\columnwidth]{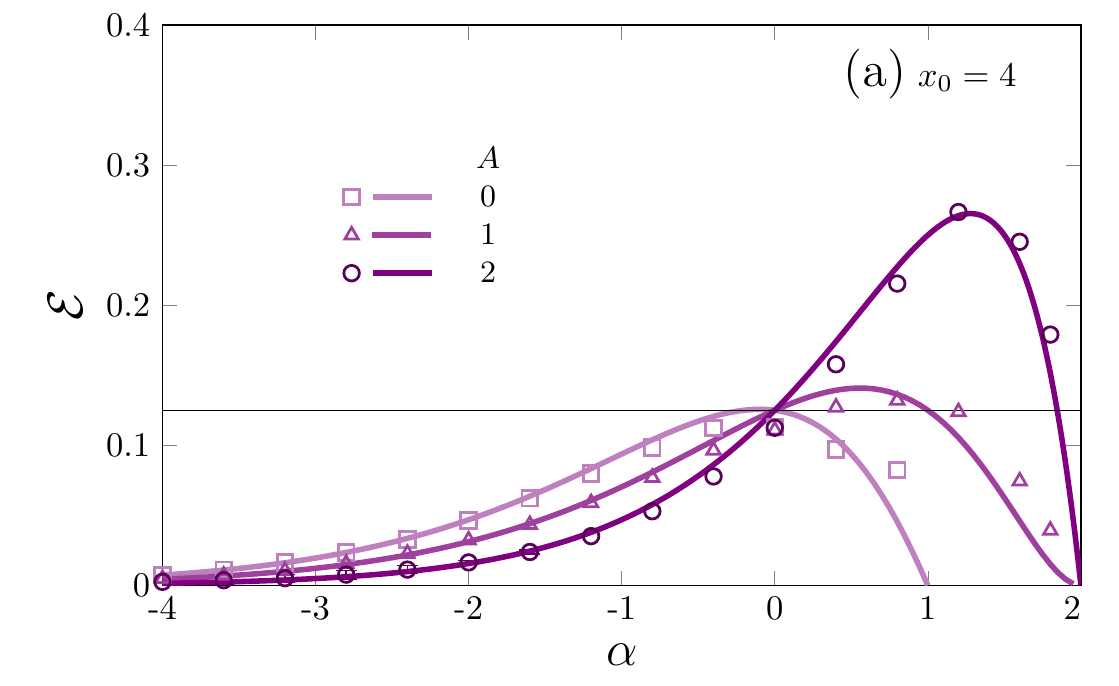}
 \includegraphics[width=0.95\columnwidth]{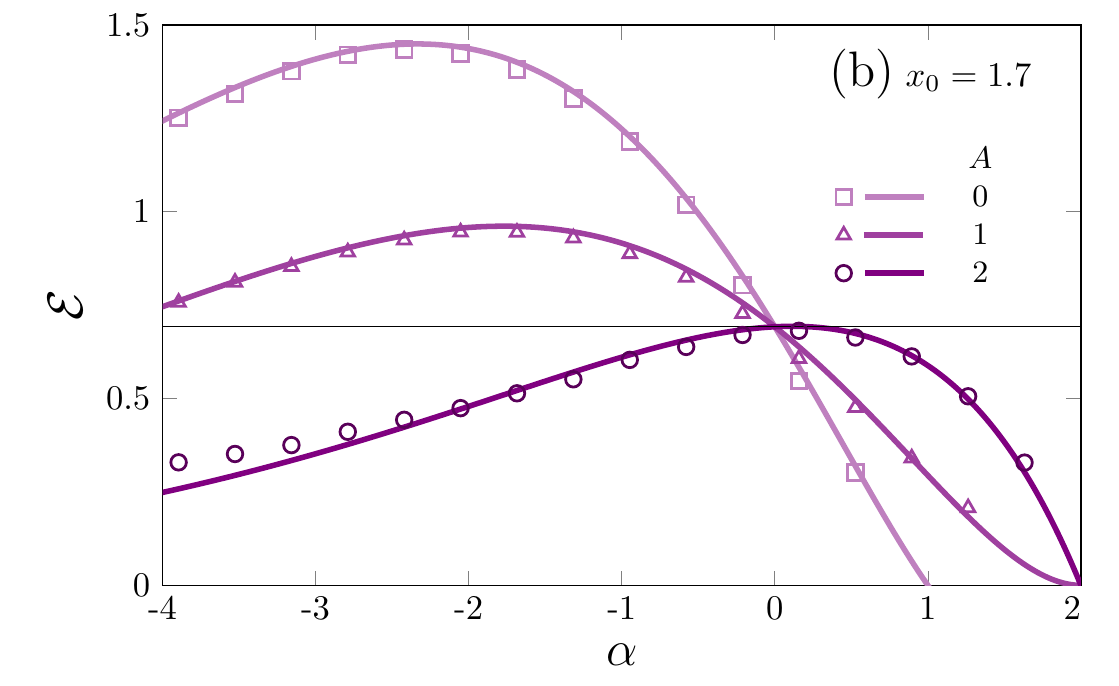}
  \includegraphics[width=0.95\columnwidth]{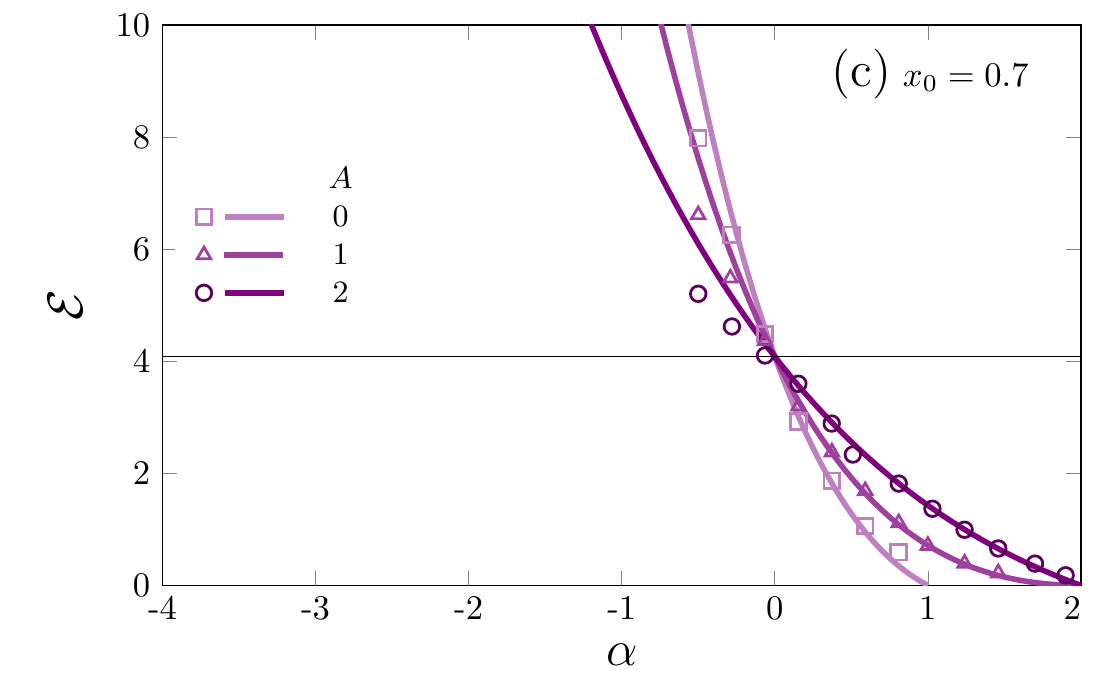}
\caption{Efficiency versus $\alpha$ for different interpretations of the HDP (values of $A$ indicated in the legend), using  $D(x)=D_0\, x^{\alpha}$, with $D_0=1$.  Each panel corresponds to a different value of $x_0$. 
The prediction given by  Eq.~(\ref{eq:ef-alfa}) is shown by solid lines, and the average over $10^5$ realizations of Eq.~(\ref{eq:ito})  with absorbing boundary at $x=0$ is represented by symbols.  
In each case, the average  over  is shown. The solid horizontal line corresponds to the homogeneous value $\mathcal{E}_H=2D_0/x_0^2$. 
} 
\label{fig:efialfa}
\end{figure}
 
This equation (\ref{eq:ef-alfa}) summarizes the effects of the heterogeneity produced by $\alpha$, under different interpretations. 
When $\alpha=0$,  the standard   efficiency for the homogeneous case, $\mathcal{E}_H  =  2 D_0 / x_0^{2}$~\cite{palyulin2017comparison}, is  recovered. 
In Fig.~\ref{fig:efialfa}, we show plots of $\mathcal{E}$  as a function of $\alpha$, for different values of $A$, generated from  Eq.~(\ref{eq:ef-alfa}), in good agreement with the results of simulations of the stochastic Eq.~(\ref{eq:ito}).
In Figs.~\ref{fig:efialfa}(a)-(b), where $x_0>1$, we note that there is an optimal value   $\alpha_{max}$, which is shifted to the right with increasing $A$. The optimal efficiency $\mathcal{E}(\alpha_{max})$, which decays with $x_0$ as expected, increases with $A$ for large enough $x_0$ (a)  but this tendency is inverted in case (b).  
For $x_0\le1$ (c),  the efficiency monotonically decreases with $\alpha$, for any $A$, diverging  for $\alpha\to -\infty$.

As a general feature, we notice  that,   for fixed $x_0$ and fixed $\alpha$, increasing $A$ enhances the efficiency when the diffusivity increases with the distance to the target ($\alpha>0$) but spoils the efficiency otherwise. Then,  a given behavior of the diffusivity around the target (ruled by $\alpha$) can be compensated by suitable correlations (ruled by $A$) in the motion of the searcher.  
%
%

\section{ Random search within the   Stratonovich scenario}
\label{sec:stratonovich}

In this section, we consider the Stratonovich framework, for general $D(x)$. 
 The Stratonovich HDP is the particular case of Eq.~(\ref{eq:HDP}), setting $A=1$, namely
\begin{eqnarray}
\frac{\partial \ }{\partial t} p(x,t|x_0) = \frac{\partial }{\partial x} \sqrt{D(x)} \frac{\partial \ }{\partial x} \sqrt{D(x)} p(x,t|x_0), \label{eq:Stratonovich} 
\end{eqnarray}
with $p( \pm\infty,t|x_0)=0$ and $p(x,0|x_0)=\delta(x-x_0)$. This equation corresponds to the stochastic process defined by Eq.~(\ref{eq:ito}) with $A=1$.

To solve the search problem, we first solve the diffusion equation (\ref{eq:Stratonovich}) without a target and  use the free solution to reproduce the boundary condition of the search problem through the method of images. 

We introduce the following change of variables  
\begin{eqnarray}
y(x) = \int_{0}^{x} \frac{1}{\sqrt{D(x')}} dx',  \label{eq:y}
 \end{eqnarray}
 which allows to rewrite Eq.~(\ref{eq:Stratonovich}) as  
 $\partial_t \Tilde{P}(y,t)  = \partial_{yy}^2 \Tilde{P}(y,t)$, 
where $\Tilde{P}(y(x),t)=\sqrt{D(x)}\,p(x,t)$.   
Its natural solution, for $y \in (-\infty,\infty)$, is $ \Tilde{P}_0(y,t) =  
\exp( -  y^2/(4t))/\sqrt{4\pi t}$. To reproduce an absorbing wall at the origin, i.e., $\Tilde{P}(y(x=0),t)=0$, we apply the method of images to the free solution with initial condition $\Tilde{P}(y(x),0)=\delta(y-y_0)$, which implies $\Tilde{P}(y,t)= \Tilde{P}_0(y-y_0,t) - \Tilde{P}_0(y+y_0,t)$. After that, we obtain 
\begin{eqnarray} \label{eq:pdfStrat}
p(x,t|x_0) & = &  \frac{\Tilde{P}(y,t) }{\sqrt{D(x)}}= \sum_{k=-1,1}k\,\frac{  
e^{- \displaystyle{\frac{ (y(x)- k\, y(x_0))^2}{4t} } } 
}{\sqrt{4\pi D(x)t}} .\;\;\;\;\;\;\;
\end{eqnarray}
This expression   works for any space-dependent diffusivity (allowing the change $y(x)$). 

\begin{figure}[b!]  
\centering
 \includegraphics[width=0.95\columnwidth]{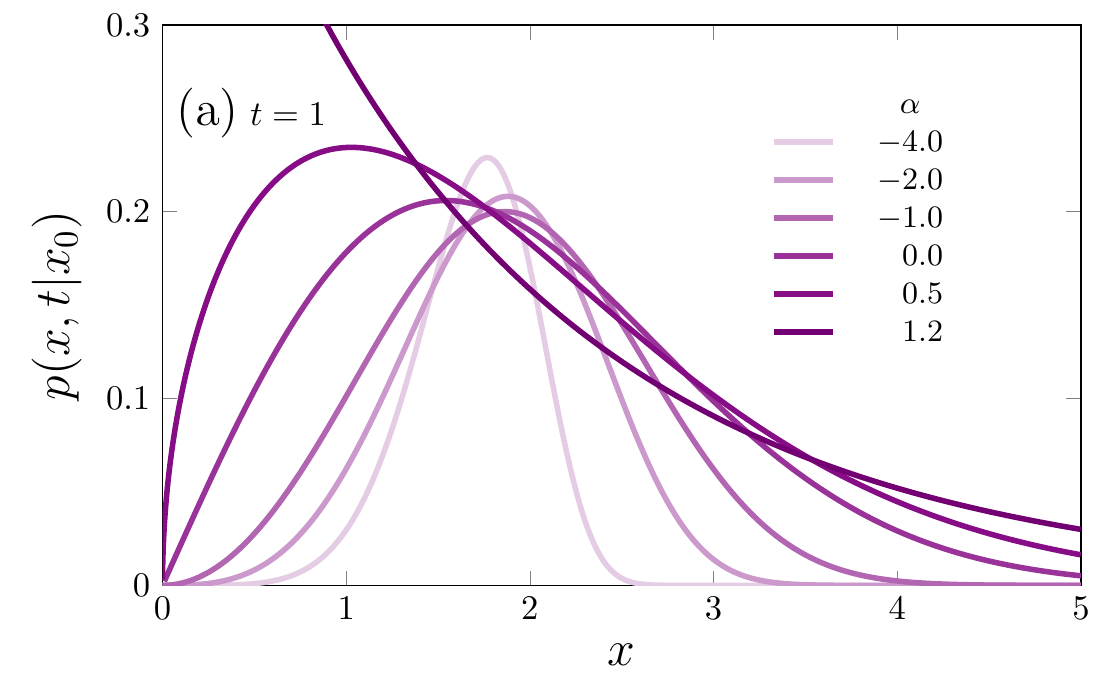}
 \includegraphics[width=0.95\columnwidth]{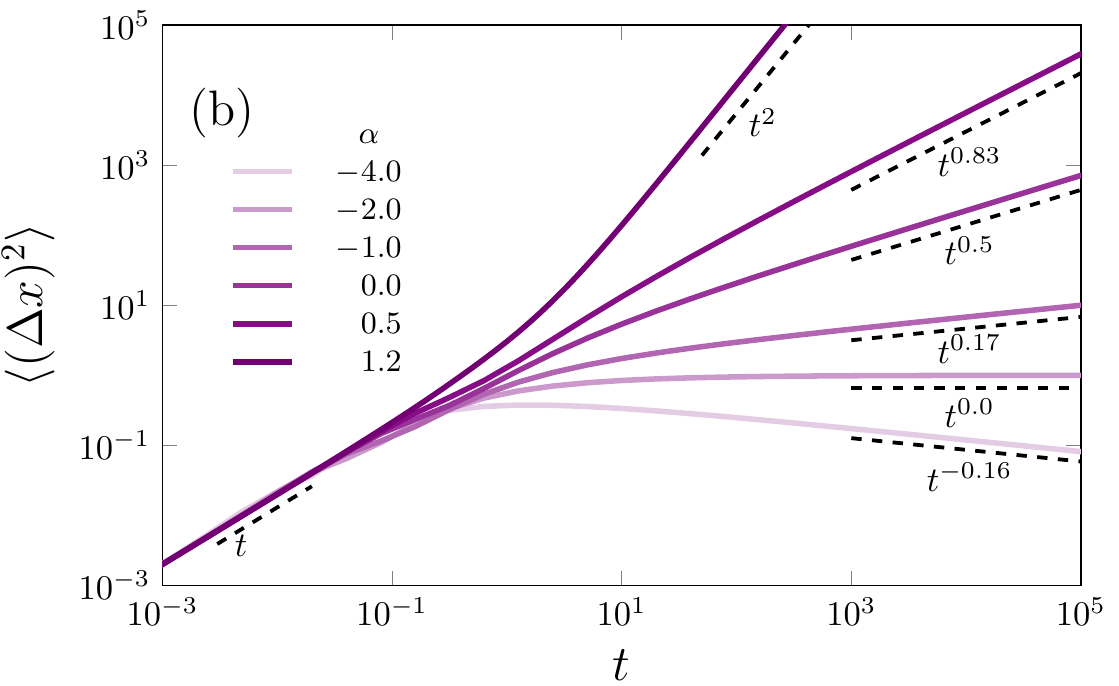}
\caption{ (a) PDF vs. $x$ at  $t=1$, and (b) MSD vs. $t$, for  $D(x)=D_0\, x^{\alpha}$, with $D_0$, and different values of $\alpha$ indicated in the legend. 
 In (a), the area under the curves smaller than unit is due to the loss of norm.  
In (b), we can observe that for short times diffusion is normal, but for long times MSD $\sim t^\frac{2+\alpha}{2(2-\alpha)}$. }
\label{fig:PDF&MSD}
\end{figure}

An  illustrative example of the PDF at a given time ($t=1$) is presented  
in Fig.~\ref{fig:PDF&MSD}   
for the power-law  diffusion coefficient given by Eq.~(\ref{eq:Dx}), with different values of $\alpha$. Notice the loss of norm,   visibly more pronounced with decreasing $\alpha$, which favors adsorption. Besides the PDF, we show the mean square displacement (MSD) versus time. 
At early times, the MSD increases linearly with time for any $\alpha$, meaning a normal diffusion spread. However, for long times, we observe an unusual behavior of the MSD, namely  $\langle (\Delta x)^2 \rangle \propto t^{\frac{2+\alpha}{2(2-\alpha)}}$. 
See derivation in Appendix~\ref{app:moments}. 
When $\alpha=-2$, the MSD becomes stationary but the PDF keeps losing norm. 
For $\alpha<2$, besides losing the norm, the PDF narrows with time as reflected by the negative exponent. 

The integration of Eq.~(\ref{eq:pdfStrat}) over $x$  yields the survival probability 
\begin{eqnarray}
Q(x_0|t) & = &  \int_0^{\infty} \frac{ e^{- \frac{(y-y(x_0))^2}{4t}} -   e^{- \frac{(y+y(x_0))^2}{4t}}}{\sqrt{4\pi t}}   dy \nonumber \\
& = & \text{erf}\left( \frac{y(x_0)}{2t^\frac{1}{2}}\right), \label{eq:survivalStrat}
\end{eqnarray}
where {\rm erf} is the error function.

The FPTD can be obtained directly from  Eq.~(\ref{eq:survivalStrat}), using Eq.~(\ref{eq:wp}),  namely
\begin{eqnarray} \label{eq:1st}
\wp(t) = \frac{y(x_0) e^{-\frac{ y(x_0)^2}{4t}}}{2\sqrt{\pi } t^{\frac{3}{2}} }\,.
\end{eqnarray}
When $D(x)$ is a power-law, we recover 
the result of Eq.~(\ref{eq:FPT}) for $A=1$.

Using $\wp(t)$ in Eq.~(\ref{eq:1st}), we compute the efficiency defined in Eq.~(\ref{eq:efficiency}) $\mathcal{E}=\langle t^{-1} \rangle=\int \wp(t) t^{-1} dt$, and through the change of variables $\xi = y_0^2/(4 t)$, we have
$\mathcal{E} 
= \frac{4 }{y_0^2} \int_0^{\infty}   \frac{  e^{-\xi}}{\sqrt{\pi } } \xi^{\frac{1}{2}}  d\xi  = \frac{2 }{y_0^2}  $. Therefore,  
\begin{eqnarray}
\mathcal{E} =
\frac{2 }{\displaystyle \bigg| \int_{0}^{x_0}  [D(x')]^{-\frac{1}{2}} dx' \bigg|^2 },  \;\;\;
\label{eq:Effgeneral}
\end{eqnarray}
valid for arbitrary diffusivity profile $D(x)$. 
Notice that the efficiency only depends on the profile within the interval $(0,x_0)$.  Moreover, notice  that, since the integrand is a function of $D(x)$ only, then the efficiency  does not depend on the particular sequence of values of the diffusivity. If we fragment the profile and shuffle the fragments~\cite{dos2020critical}, the value of the integral will be the same. This is clear if we discretize the integral in Eq.~(\ref{eq:Effgeneral}) as 
 $y_0 \simeq \frac{x_0}{N} \sum_{i=1}^N [D(x_i)]^{-1/2}$, 
 which is invariant by shuffling the values 
 of $D(x_i)$ within the integration interval. 

Moreover, to put into evidence the variations $\xi$ around a 
reference level $D_0$, we write  $D(x)=D_0[1+\xi(x)]$, 
such that $\langle \xi \rangle=0$, and $\xi>-1$ 
for the positivity of $D$. 
Under such constraints for $\{\xi_i\}$, it is 
easy to show that $ \sum_{i=1}^N [1+x_i]^{-1/2}/N\ge1$~\cite{dos2020critical}, 
then in the continuous limit $y_0\ge x_0/\sqrt{D_0}$, which implies
\begin{equation} \label{eq:discret}
  \mathcal{E} =\frac{2}{y_0^2} \le \frac{2 D_0}{x_0^2} =\mathcal{E}_H\,.
\end{equation}
This means the remarkable property that the efficiency of a heterogeneous profile is lower than that of a homogeneous profile with a level equal to the average of the heterogeneous one.

We will provide below two concrete examples:  a localized break of homogeneity, and an oscillatory profile. 
In addition we will discuss the case of a stochastic profile.

\subsection{Localized heterogeneity}

\begin{figure}[h!]  
\centering
\includegraphics[width=0.95\columnwidth]{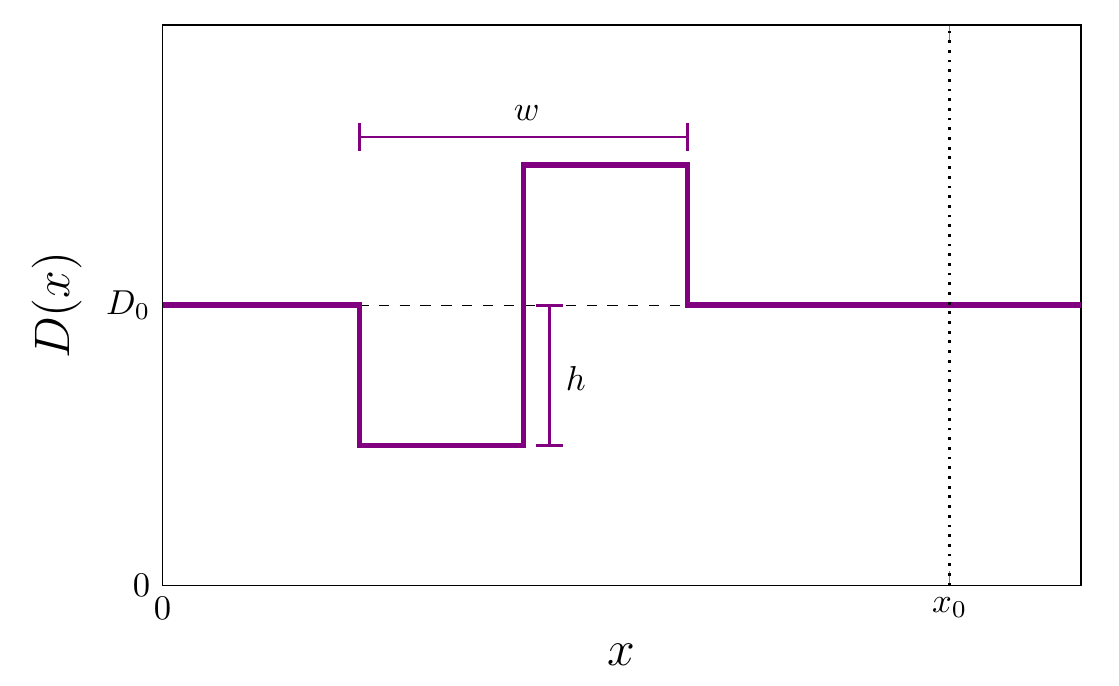}
\caption{Localized heterogeneity of width $w$ and amplitude $h$, around the level $D_0$ in dashed line. The dotted vertical line highlights the initial position.} 
\label{fig:rectangulo}
\end{figure}

We analyze a profile that presents a local perturbation of the diffusivity around the  level $D_0$. The  average diffusivity is conserved, as far as the perturbation is contained within the interval $[0,x_0]$.
The local heterogeneity  depicted in Fig.~\ref{fig:rectangulo}  has width $0\le w\le x_0$ and amplitude $0\le h\le 1$.
For this profile, Eq.~(\ref{eq:Effgeneral}) straightforwardly yields

\begin{eqnarray}
\label{eq:Erectangulo}
    \mathcal{E} 
   &=&   \frac{2 D_0 }
    { \left( x_0 -w +w \frac{ \sqrt{1+h}+\sqrt{1-h} }{2\sqrt{1-h^2} }  
    \right)^2} \;\le \; {\mathcal E}_H
\,.
\end{eqnarray}

\begin{figure}[b!]  
\centering
   \includegraphics[width=0.95\columnwidth]{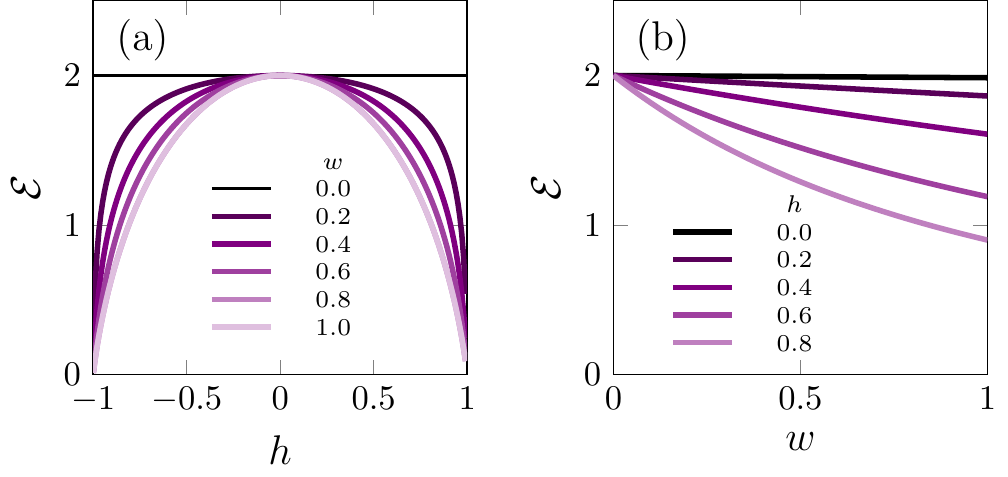}   
\caption{Efficiency for the landscape sketched in Fig.~\ref{fig:rectangulo}, varying $h$ and $w$. 
Notice that in all cases the efficiency is lower than that for the homogeneous case.  
The horizontal lines correspond to $\mathcal{E}_H $ the value for the homogeneous case with same average.}
\label{fig:E-pulse}
\end{figure}

For $w=0$ or $h=0$, the standard value $\mathcal{E}_H$ is recovered, while for increasing $w$ and $|h|$, the efficiency decays, as can be visualized in Fig.~\ref{fig:E-pulse}. 
This means that, in  a heterogeneous profile  that preserves the average, 
the search is less efficient than in an homogeneous environment with the average diffusivity.    
Let us also note, from Eq.~(\ref{eq:Effgeneral}), that a rigid shift of the pulse will not affect the efficiency, as soon as the pulse remains contained within the integration interval $[0,x_0]$. 
Also, fragmentation of the pulse into smaller ones will produce the same result, as far as the total length of up and down diffusivities is the same.
This property is related to the Stratonovich prescription and does not apply to other values of $A\neq 1$, as we will exemplify below.

\subsection{Oscillating diffusivity}
\label{sec:cosine}

\begin{figure}[b!]  
\centering
   \includegraphics[width=0.95\columnwidth]{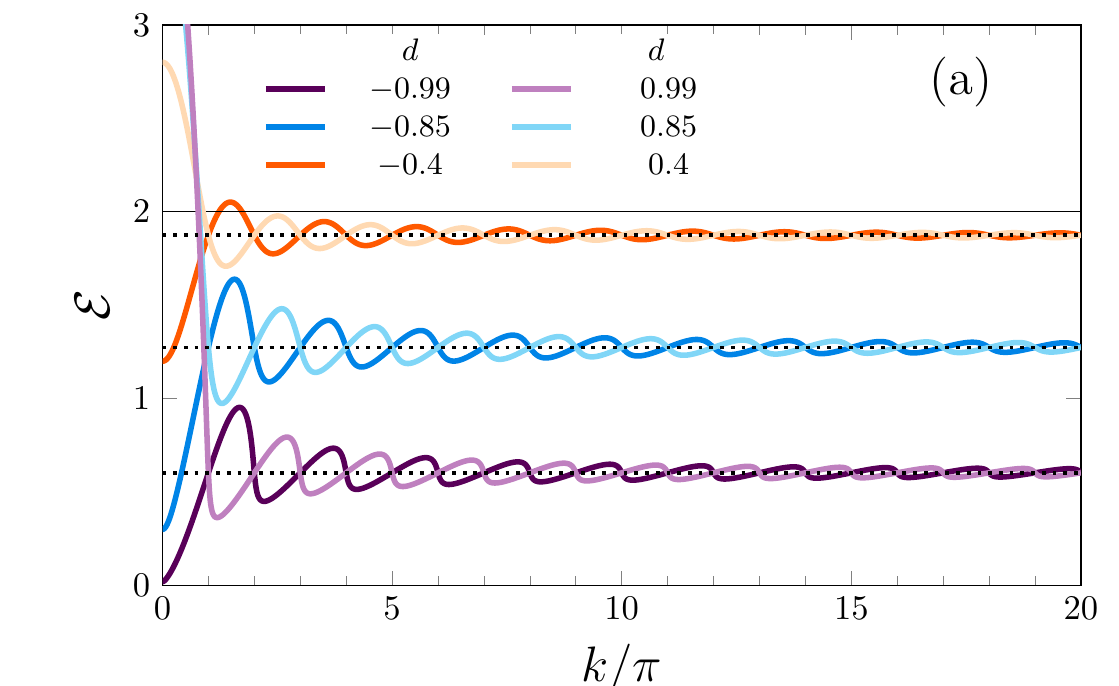} 
	\includegraphics[width=0.95\columnwidth]{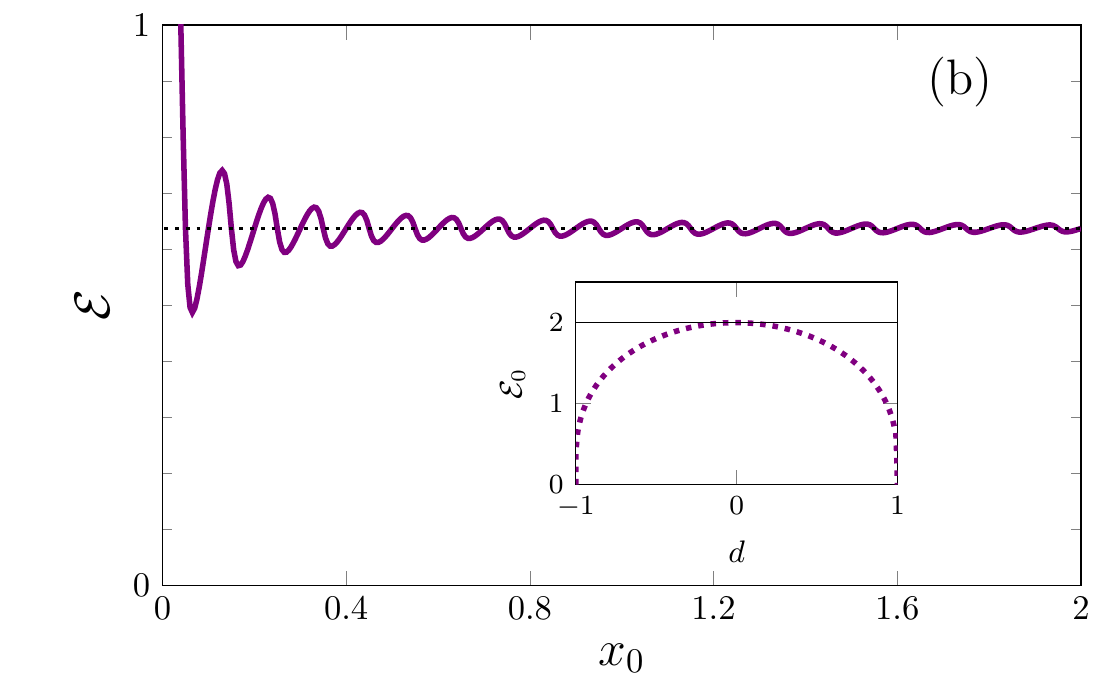}
\caption{ (a) Efficiency $\mathcal{E}$ for $D(x)= D_0[1 + d \cos(kx)]$ versus $k$, for different values of $d$. 
The dotted horizontal lines correspond to the respective short-wavelength limit given by Eq.~(\ref{eq:short}).
We set $x_0=1$ and $D_0= 1$.
In (b), we plot the short-wavelength limit $\mathcal{E}_0$ (dashed line) vs. $d$, and for comparison, the efficiency of the corresponding homogeneous case $\mathcal{E}_H$ (thin horizontal line). In the inset we plot  $\mathcal{E}/\mathcal{E}_H$ versus $x_0$, for $k=20\pi$, $D_0=1$ and $d=0.85$. }
\label{fig:effsinoidal}
\end{figure}

As a paradigm of an 
oscillatory landscape, we analyze 
the sinusoidal diffusivity kernel
\begin{equation} \label{eq:cos}
    D(x)= D_0\,[1 + d \cos (kx)]\,,  
\end{equation} 
 where     oscillations occur around the reference level $D_0$, with  $ -1\le  d\le 1$ and wavenumber $k$.
 We  integrated numerically the general expression for the efficiency, Eq.~(\ref{eq:Effgeneral}), using $D(x)$ in Eq.~(\ref{eq:cos}), showing the results in Fig.~\ref{fig:effsinoidal}.

 For a fair comparison with the homogeneous case, let us consider oscillations whose average around $D_0$ vanishes. %
 This occurs when $kx_0=N\pi$, with integer $N$, and also in the limit of very short wavelength  compared to $x_0$ (i.e., $\lambda=2\pi/k \ll x_0$). 
 For integer $N$, we obtain
  \begin{eqnarray}
    \mathcal{E}_0 \simeq
 \mathcal{E}_{ H}\frac{(1+d)\pi^2}{4 \,[{\kappa}(\frac{2d}{1+d})]^2} \;\le \;\mathcal{E}_{ H} \,,
 \label{eq:short}
\end{eqnarray}
where $\kappa(z)=\frac{\pi}{2}\,{}_2F_1 (\frac{1}{2},\frac{1}{2},1,z)$ is the complete elliptic integral of the first kind.
The short-wavelength limit    $\mathcal{E}_0$, for each $d$, is plotted in Fig.~\ref{fig:effsinoidal} by dotted horizontal lines. 
Notice that, in fact, it is attained for integer $N$ or large $k$. This limit value is independent of the introduction of a phase constant in Eq.~(\ref{eq:cos}), as can be observed when $d$  changes sign.
More importantly, Eq.~(\ref{eq:short})  is maximal at $d=0$ where it takes the value ${\mathcal{E}_H}$.  That is, the efficiency $\mathcal{E}_0$ remains below that of the homogeneous case with the same average diffusivity. 
This is a noticeable result that indicates that short-wavelength oscillations of the diffusivity spoil the efficiency of the search, which decays with increasing $d$, as represented by the dashed line in Fig.~\ref{fig:effsinoidal}(b).   In contrast, for small values of $k$, the value of the efficiency can be higher than $\mathcal{E}_H$, but this simply reflects an average diffusivity higher than $D_0$.

 \subsection{Random diffusivity}
As discussed in connection with Eq.~(\ref{eq:discret}), shuffled diffusivity profiles in the interval $(0,x_0)$  
 yield the same efficiency within Stratonovich framework. 
This leads to consider  noisy diffusivity  profiles $D(x)=D_0(1+\xi)$, around the level $D_0$,  taking uncorrelated random  values $\xi$ with a given PDF   $f(\xi)$, where  $\xi\in (-1,\infty)$, such that the average $\langle \xi \rangle=\int_{-1}^\infty \xi  f(\xi)  d\xi =0$. 
Following this idea,  Eq.~(\ref{eq:Effgeneral}) can be rewritten as 
\begin{eqnarray}\label{eq:Estoch}
\mathcal{E}=  
  \frac{ \mathcal{E}_H}{ \displaystyle \left( \int_{-1}^\infty
 [1+\xi]^{-\frac{1}{2}} f(\xi) d \xi\right)^2}\;\le \; \mathcal{E}_H,
 \label{eq:criticalsizeP}
\end{eqnarray} 
where upper bound comes from the inequality~\cite{dos2020critical} 
\begin{equation}
 \int_{-1}^\infty [1+\xi]^{-\frac{1}{2}} f(\xi) d \xi \;\ge \; 1.  
\end{equation}

Considering, that
$\xi(x) = D(x)/D_0-1$, where $x$ can be interpreted as a 
random variable that is uniform in the interval $[0,x_0]$, through the change-of-variables method, we can obtain $f(\xi)$.
For instance, in the case of $D(x)=D_0[1+d\cos( n\pi x/x_0)]$, considered in Section~\ref{sec:cosine}, it is $f(\xi)=[1-\xi^2/d^2]^{-\frac{1}{2}}/(\pi d)$, for $\xi\in(-d,d)$, 
which substituted into Eq.~(\ref{eq:Estoch}) allows to reobtain Eq.~(\ref{eq:short}).

\subsection{Comparison with other interpretations}

\begin{figure}[h!]  
\centering
   \includegraphics[width=0.95\columnwidth]{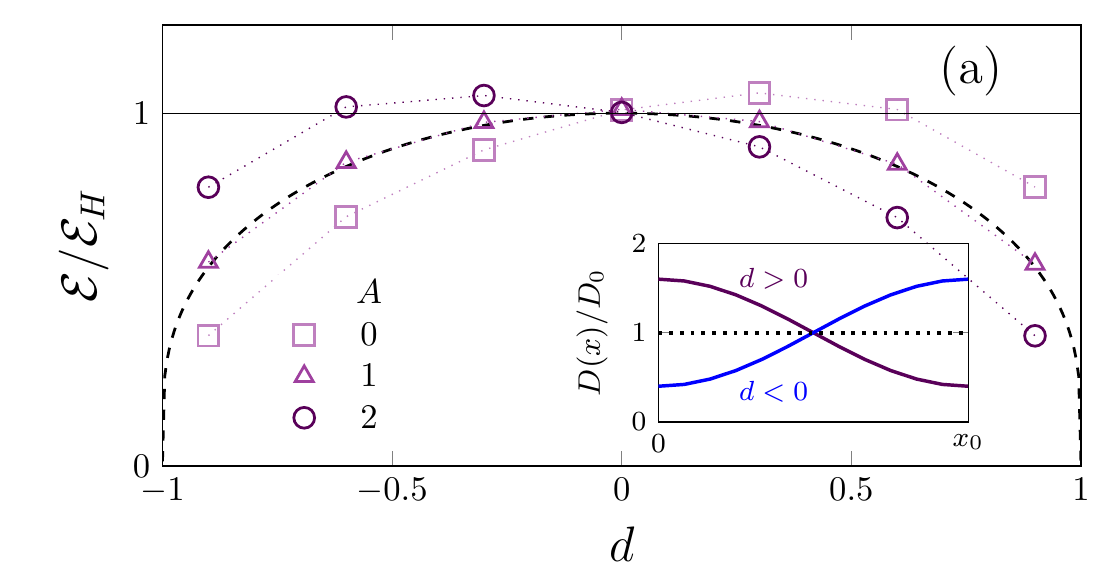} 
      \includegraphics[width=0.95\columnwidth]{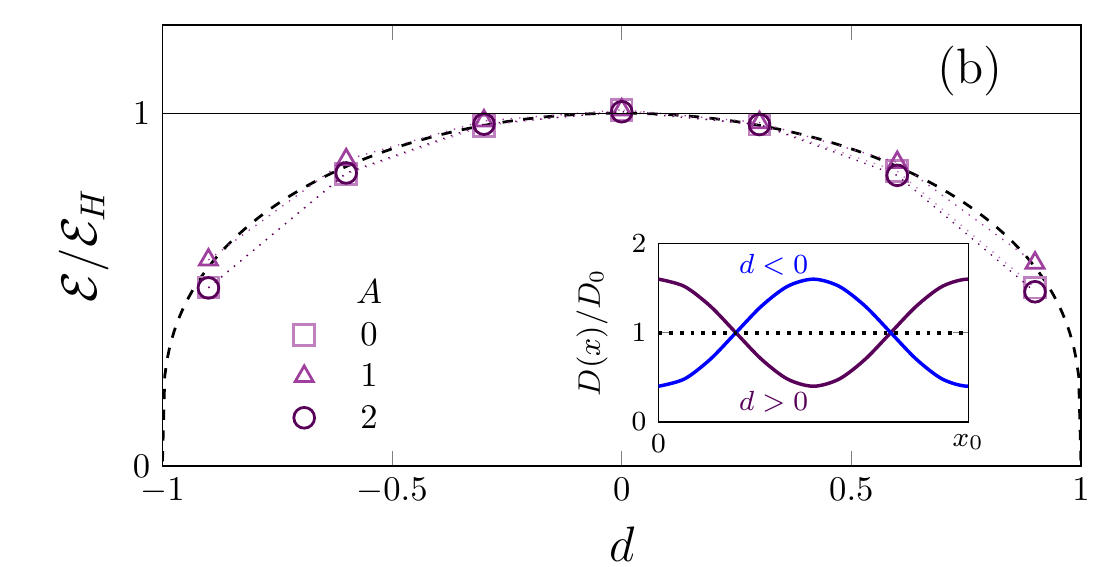}
\caption{ Relative efficiency $\mathcal{E}/\mathcal{E}_H$ 
vs. $d$ for the profile
$D(x)=D_0[1+d\cos( k x/x_0)]$ using different interpretations (values of $A$), for 
(a) $k/x_0 = \pi$ and (b) $k/x_0 = 2\pi$. 
In both cases the prediction given by Eq.~(\ref{eq:short}) for the Stratonovich case (dashed line)  and the homogeneous value (horizontal full line), both normalized by $\mathcal{E}_H$ are plotted. 
 Symbols correspond to the average over  $10^5$ trajectories of Eq.~(\ref{eq:ito}), with absorbing boundary at $x=0$. 
The diffusivity profiles are depicted in the insets. 
}
\label{fig:E012}
\end{figure}

In the precedent results of Sec.~\ref{sec:stratonovich}, we presented  results for arbitrary $D(x)$ within Stratonovich interpretation. We will perform a comparison with other interpretations using  the profile
$D(x)=D_0[1+d\cos( k x/x_0)]$ through
simulations of the stochastic differential Eq.~(\ref{eq:ito}), 
calculating and plotting the efficiency vs. $d$ in Fig.~\ref{fig:E012}. We consider cases where $k/x_0=n\pi$ with integer $n$, hence the average level is $D_0$. 

In Fig.~\ref{fig:E012}(a), where monotonic profiles are used, we can observe several features. 
For $A=1$,  the efficiency is insensitive to the ordering as proved throughout this section, then there is a symmetry of inversion  $d  \to -d$. 
However,  notice that this symmetry is broken for the other interpretations, meaning that the shape of the profile and only the distribution of values are relevant.  Moreover, we observe that when the profile increases with the distance from the target ($d<0$), the efficiency increases with $A$, while the contrary occurs for a decreasing profile ($d>0$). Actually, this is the same behavior demonstrated analytically for the power-law case analyzed in Sec.~\ref{sec:plaw}.
Finally note that while for $A=1$ the efficiency remains below that of the homogeneous profile, this can be violated for the other interpretations. 

In Fig.~\ref{fig:E012}(b), where the diffusivity profiles are not monotonic, the efficiency appears to be symmetric around $d=0$, and smaller for $A\neq1$ than for $A=1$.

\section{Final remarks}
\label{sec:final}

We have obtained the efficiency of the search problem when the medium is heterogeneous. 
For general interpretations  characterized by parameter $A$, we developed the paradigmatic case with power-law diffusivity with exponent $\alpha<2$, which embraces the cases with increasing and decreasing mobility with the distance from  the target. 
We observed that depending on the initial position of the searcher, there can be an  optimal value of $\alpha$, which depends on $A$. But a finite maximum not always occurs.  
The general feature is that, increasing $A$ favors  the search when the   diffusivity increases with the distance from the target ($\alpha>0$) and hinders the search otherwise.
Moreover, this is not unique to  the power-law shape but is determined by the monotonic character. 

The different interpretations represented by $A$ produce qualitatively similar pictures. Then, we considered the Stratonovich framework ($A=1$) that allows  deriving a closed expression of the efficiency for arbitrary forms of $D(x)$. In this case, we considered a localized perturbation and an oscillatory one, concluding that these heterogeneities spoil the efficiency of the homogeneous case with a level equal to the average one.  
It is important to note that the shape of the diffusivity profile within the search interval is not relevant but only the set of values of the profile, which determine the integral in Eq.~(\ref{eq:Effgeneral}). This is a property analogous to that found in the context of critical patch size~\cite{dos2020critical}. 
Therefore, a noisy profile with the same distribution of values yields the same results. 
However, for other interpretations other than Stratonovich one, the shape of the profile (not only the distribution of values) is relevant. 

As a perspective, it would be interesting to extend the present study to higher dimensions and confined systems. 
We want to call the attention that our results can be applied to the problem of  the first encounter between two walkers  $x(t)$ and $y(t)$, with a coupled diffusivity depending on their distance  $D(|x-y|)$. For $x_0>y_0$, $z(t)=x(t)-y(t)>0$, and for $z(t)=0$ the first encounter occurs \cite{le2020first,vot2022first}.  In such case, the efficiency measures the rate of success of the first encounter. 

{\bf Acknowledgments:} We acknowledge partial financial support by the 
Coordena\c c\~ao de Aperfei\c coamento de Pessoal de N\'{\i}vel Superior
 - Brazil (CAPES) - Finance Code 001. C.A. also acknowledges partial support by 
 Conselho Nacional de Desenvolvimento Cient\'{\i}fico e Tecnol\'ogico (CNPq), 
and Funda\c c\~ao de Amparo \`a Pesquisa do Estado do Rio de Janeiro (FAPERJ).

\appendix

\section{Solving Eq.~(\ref{eq:surveq})}
\label{app:solution}
First, we introduce the new function
\begin{eqnarray}
\Tilde{q}(x_0,s) =  \Tilde{Q}(x_0,s) - \frac{1}{s}, \label{eq:survivalLaplacespace}
\end{eqnarray}
into Eq.~(\ref{eq:surveq}), obtaining
\begin{eqnarray}
&&\frac{\partial^2 \ }{\partial x_0^2} \Tilde{q}(x_0,s)  +  \left(1-\frac{A}{2}\right) \frac{\alpha }{x_0} \frac{\partial }{\partial x_0}  \Tilde{q}(x_0,s) 
\nonumber \\&&
- \frac{s}{D_0  x_0^{\alpha}} \Tilde{q}(x_0,s) = 0.  \label{eq:newQ}
\end{eqnarray}
Using the change of variables 
\begin{eqnarray} \nonumber
z &=&\sqrt{x_0 }\\ \nonumber
\Tilde{q}(x_0,s) &=& z^{\nu}\Tilde{w}(z,s),  
\end{eqnarray} 
in Eq. (\ref{eq:newQ}), we get
\begin{eqnarray}
&&\frac{\partial^2 \ }{\partial z^2} \Tilde{w}(z,s)  + \frac{\mathcal{A}}{z} \frac{\partial }{\partial z}  \Tilde{w}(z,s) \nonumber \\
&&+ \left(\frac{\mathcal{B}}{z^2}- \left(2\frac{\sqrt{s}}{\sqrt{D_0}  z^{\alpha-1}}\right)^2 \right) \Tilde{w}(z,s) = 0, \label{eq:lommel}
\end{eqnarray}
where $\mathcal{A} =   2\nu -1 + \left(1-A/2 \right) 2 \alpha$ and 
$\mathcal{B}  =   \nu \left( \nu - 2 + 2 \alpha \left( 1 - A/2 \right) \right)$. 
Equation~(\ref{eq:lommel}) can be identified with a Lommel-type  equation~\cite{gradshteyn2007table}, which admits the solution
\begin{eqnarray}
\Tilde{w}(z,s) = c_1 z^{\beta} K_{b} \left[ a z^{ 2- \alpha} \right] + c_2 z^{\beta} I_{b} \left[ a z^{ 2- \alpha} \right], \label{eq:Wfunction1}
\end{eqnarray} 
where $K_b(\cdots)$ and $I_b(\cdots)$ are the modified Bessel functions~\cite{abramowitz1965handbook}, 
\begin{eqnarray}
a & = & \frac{2}{2-\alpha }\sqrt{\frac{s}{D_0}}, \\
\beta &=&  1 - \nu  - \alpha \left(1-A/2 \right), \label{eq:gbeta1} \\ 
b & = &  \pm [1-\alpha(1-A/2)]/(2-\alpha), 
\label{eq:gbeta2}
\end{eqnarray}
here the $\pm$ can be ignored since $K_b(z)=K_{-b}(z)$.
To ensure the convergence of the solution $\Tilde{w}(z,s)$, for large $z$,   we must set $c_2=0$ into the Eq.~(\ref{eq:Wfunction1}). 
Therefore, according Eq.~(\ref{eq:y}) $\Tilde{q}(x_0,s) = z^{\nu}w(z,s)$, we obtain 
\begin{eqnarray}
\Tilde{q}(x_0,s) = c_1\, x_0^{\frac{1}{2}(1 -  \alpha [1-A/2])} K_b \left( \frac{2\,x_0^{\frac{2-\alpha}{2}}}{2-\alpha} \sqrt{\frac{s}{D_0}}  \right).
\end{eqnarray}

The probability of survival in Laplace space (see Eq.~(\ref{eq:survivalLaplacespace})) is given by
\begin{eqnarray} \nonumber 
\Tilde{Q}(x_0,s) = c_1 \,x_0^{\frac{1}{2}(1 -  \alpha [1-A/2])} K_b \left( \frac{2\,x_0^{\frac{2-\alpha}{2}}}{2-\alpha} \sqrt{\frac{s}{D_0}}  \right) +  \frac{1}{s},
\end{eqnarray}
where the coefficient $c_1$ is obtained from the boundary condition $\Tilde{Q}(x_0=x_a,s)=0$, where $x_a$ is the target position.
 Then 
\begin{eqnarray} \nonumber 
\Tilde{Q}(x_0,s) = \frac{1}{s}\left[  1-  \,\frac{x_0^{\frac{1}{2}(1 -  \alpha [1-A/2])} K_b \left( \frac{2\,x_0^{\frac{2-\alpha}{2}}}{2-\alpha} \sqrt{\frac{s}{D_0}}  \right)}{x_a^{\frac{1}{2}(1 -  \alpha [1-A/2])} K_b \left( \frac{2\,x_a^{\frac{2-\alpha}{2}}}{2-\alpha} \sqrt{\frac{s}{D_0}}  \right)}\right],
\end{eqnarray}
taking the limit $x_a \to 0$ in the part of function that contain the $x_a$ parameter, we obtain
\begin{eqnarray}
\mathcal{I} & = & \lim_{x_a \to 0 } x_a^{\frac{1}{2}(1 -  \alpha [1-A/2])} K_b \left( \frac{2\,x_a^{\frac{2-\alpha}{2}}}{2-\alpha} \sqrt{\frac{s}{D_0}}  \right) \nonumber \\
&\simeq & \lim_{x_a \to 0 }  x_a^{\frac{1}{2}(1 -  \alpha [1-A/2])}  \frac{\Gamma[b]}{2^{1-b}}\left( \frac{2\,x_a^{\frac{2-\alpha}{2}}}{2-\alpha} \sqrt{\frac{s}{D_0}} \right)^{-b} \nonumber \\
& = & \lim_{x_a \to 0 }   x_a^{\frac{2-\alpha}{2}\left(\frac{1 -  \alpha [1-A/2]}{2-\alpha} - b \right)}  \frac{\Gamma[b]}{2^{1-b}}\left( \frac{2}{2-\alpha} \sqrt{\frac{s}{D_0}} \right)^{-b} \nonumber \\
& = & \frac{\Gamma[b]}{2^{1-b}}\left( \frac{2}{2-\alpha} \sqrt{\frac{s}{D_0}} \right)^{-b}\,,
\end{eqnarray}
which is non null only for $ b= [1-\alpha(1-A/2)]/(2-\alpha)$.  
 
Therefore,  
\begin{eqnarray}  
&&\Tilde{Q}(x_0,s) =  \label{eq:Qfinal}  \\
&=& \frac{ 1 }{s} \left[1 -  
 \frac{2  }{   \Gamma[b]}\left( \frac{x_0^\frac{2-\alpha }{2}}{2-\alpha} \sqrt{\frac{s}{D_0}} \right)^{b}   K_b \left( \frac{2\,x_0^{\frac{2-\alpha}{2}}}{2-\alpha} \sqrt{\frac{s}{D_0}}  \right)  \right]. \nonumber
\end{eqnarray}
The FPTD in Eq.~(\ref{eq:Qfinal})  is normalized only for $b>0$, then
\begin{equation}
    1-\alpha(1-A/2)    \ge 0.   \label{eq:constraint}
\end{equation}

\section{Mean square displacement (MSD) } \label{app:moments}

We calculate the second moment, which determines the asymptotic long-time limit presented in Fig.~\ref{fig:PDF&MSD}. To do that we perform the average using Eq.~(\ref{eq:pdfStrat})  as
\begin{eqnarray} \nonumber
&&\mathcal{L}\left\{\langle x^{2} \rangle \right\} = \int_{0}^{\infty} x^2 p(x,t) dx \\  = &&  \int_0^{\infty} \frac{x^{2-\frac{\alpha}{2}}}{\sqrt{4 s}} \left(  e^{- \sqrt{s}|y(x)-y(x_0)|} -   e^{- \sqrt{s}|y(x)+y(x_0)|}  \right) dx \nonumber \\
 =&&   e^{-\sqrt{s}\, y(x_0)}  \int_0^{x_0} \frac{x^{2-\frac{\alpha}{2}}}{\sqrt{4 s}}  e^{\sqrt{s}\, y(x)}  dx \nonumber \\
&& + 
e^{ \sqrt{s}\,y(x_0)} \int_{x_0}^{\infty} \frac{x^{2-\frac{\alpha}{2}}}{\sqrt{4 s}}  e^{-\sqrt{s}\,y(x)}  dx \nonumber \\
&& -   e^{-\sqrt{s}\,y(x_0)}   \int_{0}^{\infty} \frac{x^{2-\frac{\alpha}{2}}}{\sqrt{4 s}}  e^{-\sqrt{s}\,y(x)}dx   \nonumber \\
 = &&  \frac{\sinh( {\sqrt{s}\,y(x_0)})  }{\sqrt{ s}} \int_{x_0}^{\infty}  x^{2-\frac{\alpha}{2}}   e^{-\sqrt{s}\,y(x)}   dx\,,\\ \nonumber
 &&+ 
   \frac{e^{-\sqrt{s}\,y(x_0)}}{\sqrt{ s}} \int_{0}^{x_0}  x^{2-\frac{\alpha}{2}}  
   \sinh( {\sqrt{s}\,y(x)}) \,dx\,,
\end{eqnarray} 
where $y$ was defined in Eq.~(\ref{eq:y}).
Defining $z = \sqrt{s}\,y$ we get
\begin{eqnarray}
z(x)= 2 s^\frac{1}{2} x^{1-\frac{\alpha}{2}}/(2-\alpha)\,,
\end{eqnarray}
which implies $x = c_{\alpha} \left(  z/\sqrt{s}\right)^{\frac{2}{2-\alpha}}$
with $c_{\alpha} = \left( \frac{2-\alpha}{2} \right)^{\frac{2}{2-\alpha}}$.

\begin{eqnarray} \nonumber
  \mathcal{L}\left\{\langle x^{2} \rangle \right\}   &=&     c_{\alpha}^2\frac{\sinh( {\sqrt{s}\,y(x_0)})}{s^  \frac{4-\alpha}{2-\alpha}}    \int_{\sqrt{s}\,y_0}^{\infty}    z^\frac{4}{2-\alpha}    \, e^{-z}  dz \,.  \nonumber \\ 
& +& 
 c_\alpha^2
\frac{e^{-\sqrt{s}\,y(x_0)}   }{s^  \frac{4-\alpha}{2-\alpha}} 
\int_{0}^{ \sqrt{s} y_0} z^\frac{4}{2-\alpha}     \sinh( {z}) \,dz\,,\label{eq:Lx2}
\end{eqnarray}
where $z(x_0)=\sqrt{s} y(x_0)\equiv \sqrt{s} y_0$.

For large $t$, i.e., $s\sim 0$, the first term in Eq.~(\ref{eq:Lx2}) dominates and we obtain
\begin{eqnarray}
 \mathcal{L}\left\{\langle x^{2} \rangle \right\} \bigg|_{s\sim 0 } &\simeq&    \frac{c_{\alpha}^2\,y_0 }{(\sqrt{s})^\frac{6-\alpha}{2-\alpha}} \int_{0}^{\infty}    z^{\frac{4}{2-\alpha}}   e^{-z} dz\nonumber \\
 & \simeq  &  \frac{2y_0c_{\alpha}^{1+\frac{\alpha}{2}}  \Gamma\left[\frac{4}{2-\alpha}\right]}{(\sqrt{s})^\frac{6-\alpha}{2-\alpha} }\,. 
\end{eqnarray}

Therefore, we obtain the asymptotic behavior
\begin{eqnarray}
 \langle x^{2} \rangle 
&& \simeq     
\frac{2 y_0 
c_\alpha^{1+\frac{\alpha}{2}} 
\Gamma\left[\frac{4}{2-\alpha}\right]}
{\Gamma\left[\frac{1}{2}\frac{6-\alpha}{2-\alpha}\right]}
\;  t^\frac{2+\alpha}{2(2-\alpha)}
\sim t^\frac{2+\alpha}{2(2-\alpha)} \,.
\end{eqnarray}

Notice that for $\alpha=0$, normal diffusion is not obtained, due to the presence of the absorbing wall~\cite{redner2001guide}.

\end{document}